\documentclass[letterpaper, 10 pt, conference]{IEEEtran}  % Comment this line out
\IEEEoverridecommandlockouts                              % This command is only
\usepackage{graphicx}
\usepackage{multirow}
\usepackage{array}
\usepackage[hyphens]{url}
\usepackage{makecell}
\usepackage{color}
\usepackage{placeins}
\usepackage[bottom]{footmisc}
\title{
Optimization of Lightweight Malware Detection Models For AIoT Devices%*
}

\author{
    \IEEEauthorblockN{Felicia Lo$^{1}$, Shin-Ming Cheng$^{3}$ ~\IEEEmembership{Member, ~IEEE}}
    \IEEEauthorblockA{National Taiwan University of Science and Technology, \\ Department of Computer Science and Information Engineering, \\ Email: $^{1}$b10815009@gapps.ntust.edu.tw, $^{3}$smcheng@mail.ntust.edu.tw} \\
    \and
    \IEEEauthorblockN{Rafael Kaliski$^{2}$ ~\IEEEmembership{Member, ~IEEE}}
    \IEEEauthorblockA{$^{2}$National Sun Yat-sen University, \\ Department of Computer Science and Engineering, \\ Email: $^{2}$rkaliski@ieee.org}
\thanks{Rafael Kaliski (corresponding author) is thankful for funding from Taiwan’s 
    National Science and Technology Council (NSTC), grants: NSTC 108-2218-E-011-036-MY3 and NSTC 112-2221-E-110-023-}%
\thanks{©2023 IEEE. Personal use of this material is permitted. Permission from IEEE must be
obtained for all other uses, in any current or future media, including
reprinting/republishing this material for advertising or promotional purposes, creating new
collective works, for resale or redistribution to servers or lists, or reuse of any copyrighted
component of this work in other works.}%
}

\begin{document}

\maketitle
\thispagestyle{empty}
\pagestyle{empty}

%%%%%%%%%%%%%%%%%%%%%%%%%%%%%%%%%%%%%%%%%%%%%%%%%%%%%%%%%%%%%%%%%%%%%%%%%%%%%%%%
\begin{abstract}

Malware intrusion is problematic for Internet of Things (IoT) and Artificial Intelligence of Things (AIoT) devices as they often reside in an ecosystem of connected devices, such as a smart home. If any devices are infected, the whole ecosystem can be compromised. Although various Machine Learning (ML) models are deployed to detect malware and network intrusion, generally speaking, robust high-accuracy models tend to require resources not found in all IoT devices, compared to less robust models defined by weak learners. In order to combat this issue, Fadhilla  \cite{lightweight} proposed a meta-learner ensemble model comprised of less robust prediction results inherent with weak learner ML models to produce a highly robust meta-learning ensemble model. The main problem with the prior research is that it cannot be deployed in low-end AIoT devices due to the limited resources comprising processing power, storage, and memory (the required libraries quickly exhaust low-end AIoT devices' resources.) Hence, this research aims to optimize the proposed super learner meta-learning ensemble  model\cite{lightweight} to make it viable for low-end AIoT devices. We show the library and ML model memory requirements associated with each optimization stage and emphasize that optimization of current ML models is necessitated for low-end AIoT devices. Our results demonstrate that we can obtain similar accuracy and False Positive Rate (FPR) metrics from high-end AIoT devices running the derived ML model, with a lower inference duration and smaller memory footprint.

\end{abstract}


%%%%%%%%%%%%%%%%%%%%%%%%%%%%%%%%%%%%%%%%%%%%%%%%%%%%%%%%%%%%%%%%%%%%%%%%%%%%%%%%
\section{Introduction}
    \label{sec:Introduction}
    For the Internet of Things (IoT), Malware intrusion remains a significant problem due to the risk posed to the local connected ecosystem; private and sensitive information can be stolen, and the infected device can act as an agent for an intruder and assist in executing their nefarious plans, such as acting as part of a BotNet and recruiting other connected devices. Compared to IoT, this problem becomes more pronounced in Artificial Intelligence of Things (AIoT) devices because AIoT devices can connect to, and often coordinate with, other local devices in the ecosystem. Furthermore, both IoT and AIoT devices are typically specific-purpose-built devices that, once deployed in the field, cannot easily have additional features added. Hence, detecting malware before it gets a foothold in an IoT device is essential to the success of IoT.

    In order to detect malware intrusions, various machine learning (ML) models are typically deployed, such as Deep Neural Networks (DNNs), Convolutional Neural Networks (CNNs), Long-short term memories (LSTM) \cite{FL_CYBER}. The ML models learn the characteristics of the previously detected malware and then use this knowledge to detect future malware. \cite{malware_detection_method} categorizes the malware detection ML models into two groups based on the detection method. The first method detects the execution characteristics, while the second detects the network traffic characteristics of the malware. Although robust deep learning models can detect the malware, higher computational power is needed for their deployment. On the other hand, if lower computational power weak learners are used, the detector may not be robust enough, i.e., it may suffer from low accuracy and a high False Positive Rate (FPR).

    To enable malware detection on IoT devices while addressing the accuracy and FPR issues associated with weak learners, \cite{lightweight} proposed a meta-learner lightweight ensemble model to detect malware via its network characteristics. Meta-learning ensemble models combine the prediction results of the multiple "constituent" weak learners to effectively boost malware detection accuracy and lower FPR. Although ensemble meta-learners are robust, they are unfortunately not necessarily compatible with low-end AIoT devices due to the memory requirements of said ML models and the associated libraries. Hence, this research aims to convert the highest performing proposed model (super learner) from \cite{lightweight} while optimizing it for low-end devices.
    Our primary contributions are: 
    \begin{itemize}
        \item We analyze the super learner ensemble model and determine that inefficient ML models result in a non-necessary increase in model inference duration and size.
        \item We present a methodology to convert a Personal Computer-based ensemble ML model to a performance metric equivalent model that will fit in a resource-limited low-end IoT device, yet obtain a lower inference duration. 
        \item We empirically analyze the number of Multi-Layer Perceptron (MLP) hidden nodes and how the Random Forest (RF) size affects the ML model's inference duration and accuracy.
        %; i.e., we reduce the resource requirements and remove Python ML-specific libraries while simplifying the ML model.
    \end{itemize}

    The remainder of this paper is organized as follows: In section \ref{sec:Related_Work}, we discuss the background and related work. In section \ref{sec:System_Model}, our AIoT model is discussed. %, while in section \ref{sec:AIoT_Limitations}, the \textcolor{red}{problem statement} of is presented. 
    Section \ref{sec:Optimization} discusses the optimization methodologies employed. Then, in section \ref{sec:Simulation}, the simulation results are presented, while the results are discussed in section \ref{sec:Results_Discussion}. Finally, we present our conclusion and future work in section \ref{sec:Conclusion}.

\section{Related Work}
    \label{sec:Related_Work}
    %Based on the IoT Analytics Spring 2022 report \cite{IoT_Analytics_Rep}, there are around 12.2 billion IoT device by the end of 2021 and they projected that by 2025, there would be more than 27 billion connected IoT devices. Due to the building of ecosystem of IoT devices, the increasing number of IoT devices call for the need of better network security model.

    This research uses the statistic-based method of the network traffic analysis \cite{malware_detection_method} generated using Zeek~\cite{Garcia} to detect malware. 
    After obtaining the statistical data, ML models are trained and used in AIoT devices for malware detection. Due to the ever-changing nature of malware, the ML models are constantly being updated. Yet, due to the memory limitation of AIoT devices, the training of the ML models is commonly done in a cloud server and then deployed to AIoT devices. Each AIoT device has a complete ML model for detection and must retrieve updated models at regular intervals.
    
    Numerous ML models have been deployed for detecting malware, ranging from traditional "weak learning" ML models, such as Decision Tree (DT), RF, and Naive Bayes (NB), up to more advanced ML models, such as DNNs, CNNs, and LSTMs. Although weak learners require fewer resources (processing power and memory), the main issue in using traditional ML models is their inability to detect a diverse set of malware. On the other hand, although advanced ML models can detect a diverse set of malware, AIoT devices usually lack the resources required to use them. In order to address this issue, ensemble techniques may be used to combine the prediction results of weak ML models to achieve a higher robustness compared to any traditional weak ML model while requiring fewer resources compared to advanced ML techniques. %\textcolor{red}{add citations}

    The most common learners in supervised ensemble classification techniques include Bagging, AdaBoost, RF, Random Subspace, and Gradient Boosting \cite{Ensemble_Learning_Survey}. Another emerging supervised ML technique uses ensemble classification and employs meta-learners to make better predictions from the ensemble learners' results. Meta-learning ensemble learners are composed of multiple weak learner layers and a final meta-learner layer. The lower layers initially process the input data via multiple ML models; in general, ensemble learners work best when the weak learners produce uncorrelated results. 
    Then, the prediction results of the lower layers are combined by the meta-learner, which is also a ML model. The commonly used models in this group include super learner ensemble, subsemble learner ensemble, and sequential ensemble. 

    The super learner model \cite{superlearner}, based on the stacking ensemble, stacks the results from each prediction layer and uses them in the next layer. The subsemble learner \cite{subsemble_learner}, on the other hand, uses bagging and k-fold cross-validation to obtain prediction results. With subsemble learners, the data is initially divided into groups. In each layer, each group uses k-fold cross-validation to train the ML models in that group of the layer. The prediction results from each group are then stacked together and used as input by the next layer. Finally, the sequential ensemble is a meta-learner ensemble model where each layer is constructed independently, i.e., one layer can be based on a super learner. In contrast, another layer can be based on a subsemble learner.

    Fadhilla \cite{lightweight} used scikit\cite{scikit-learn} and ML-Ensemble (MLENS) \cite{mlens} to build the super learner, the subsemble learner, and sequential ensembles to detect malware in a resource-efficient manner (each of the meta-learner ensemble models tested consists of three distinct layers; the base layer, the intermediate layer, and the meta learner layer.) The super learner and the subsemble learner models consist of the same weak learners in each of their layers, as shown in Table~\ref{tbl:layer constituent}. At the same time, the sequential ensemble is built by using a subsemble learner as the base layer and a super learner as the intermediate layer. All learners use a MLP as the meta layer. Each model was tested on a Raspberry Pi 4 device. The results show that the meta-learner ensemble models are robust in malware detection. Furthermore, the super learner model was determined to provide the best overall performance and inference duration.

    \begin{table}[b]
    \centering
    \caption{Weak learners comprising each layer of the super learner and subsemble learner models, as proposed by \cite{lightweight}}
    \label{tbl:layer constituent}
    %\resizebox{\columnwidth}{!}{%
    \begin{tabular}{|c|c|}
    \hline
    \textbf{Layer}          & \textbf{Weak Learner Constituent}        \\ \hline
    Base          & Random Forest and Logistic Regression    \\ \hline
    Intermediate  & Decision Tree and Multi Layer Perceptron \\ \hline
    Meta          & Multi Layer Perceptron                   \\ \hline
    \end{tabular}%
    %}
    \end{table}

    Although \cite{lightweight} demonstrated that their meta-learner ensemble model could be robust enough to be used in AIoT devices, using the MLENS library and Python limits the feasibility of deploying a meta-learner in low-end AIoT devices. This limitation is problematic as many IoT and AIoT devices are deployed with few additional resources, i.e., they lack the resources equivalent to the Raspberry Pi 4 testbed. Hence, this research aims to re-code the meta-learner ensemble models to be suitable for low-end AIoT devices while obtaining equivalent performance metrics.

\section{AIoT Model}
    \label{sec:System_Model}
    We define AIoT as Graphics Processing Unit free IoT devices running Artificial Intelligence (AI) applications.  We focus on low-end IoT devices with insufficient resources for either Python or MLENS deployment.  The overhead associated with Python easily precludes low-end devices such as the ESP8266 or Raspberry Pi Pico (a sampling of the IoT device microcontrollers, their clock speeds, and memories\footnote{ROMs are non-volatile writable storage.} %, while RAMs are volatile sotrage
    used are shown in Table~\ref{tbl:AIoT devices}.) In addition, as low-end IoT devices do not have a conventional operating system, MLENS cannot easily be ported. 
    
    \begin{table}[b]
    \centering
    \caption{Examples of WiFi-enabled AIoT micro-controllers. Low-end devices include ESP8266, ESP32, and Raspberry Pi Pico, while the Raspberry Pi Model B is considered mid to high-end.} %/ AIoT microcontroller.}     %mid/high-end AIoT microcontroller
    \label{tbl:AIoT devices} %include ESP8266, ESP32, and Raspberry Pi Pico.  
    %\resizebox{\columnwidth}{!}{%
    \begin{tabular}{|c|ccc|cc|}
    \hline
    \multirow{3}{*}{\textbf{Device}} &
    \multicolumn{3}{c|}{\textbf{Internal}} &
    \multicolumn{2}{c|}{\textbf{External}} \\ \cline{2-6} 
    &
    \multicolumn{1}{c|}{\makecell{\textbf{Clock} \\ \textbf{Speed}}} &
    \multicolumn{1}{c|}{\textbf{ROM}} &
    \textbf{RAM} &
    \multicolumn{1}{c|}{\textbf{ROM}} &
    \textbf{RAM} \\ \hline
    ESP8266  & \multicolumn{1}{c|}{80 MHz}  & \multicolumn{1}{c|}{384 kB} & 160 kB & \multicolumn{1}{c|}{16 MB} & 4 MB \\ \hline
    ESP32    & \multicolumn{1}{c|}{240 MHz} & \multicolumn{1}{c|}{448 kB} & 520 kB & \multicolumn{1}{c|}{16 MB} & 4 MB \\ \hline
    \begin{tabular}[c]{@{}c@{}}Raspberry \\ Pi Pico\end{tabular} & \multicolumn{1}{c|}{133 MHz} & \multicolumn{1}{c|}{2 MB}   & 264 kB & \multicolumn{1}{c|}{-}     & -    \\ \hline
    \begin{tabular}[c]{@{}c@{}}Raspberry \\ Pi Model B\end{tabular} &
    \multicolumn{1}{c|}{1500   MHz} &
    \multicolumn{1}{c|}{-} &
    $\le$8   GB & %1/2/4/8   GB &
    \multicolumn{1}{c|}{\begin{tabular}[c]{@{}c@{}}64 GB \\ SD Card\end{tabular}} &
    - \\ \hline
    \end{tabular}%
    %}
    \end{table}

    In general, due to the smaller memory size, there is often limited library support when using low-end microcontrollers. For microcontroller applications requiring Python, micropython \cite{micropython} is commonly used. While some libraries are rewritten to better fit into a microcontroller framework, such as ulab, the micro version of numpy, MicroPython has limited library support. Two possible solutions are that we can either minimize library usage or convert the model to C. The C model can be executed directly on low-end microcontrollers.
    
%\section{AIoT Limitations}
%    \label{sec:AIoT_Limitations}
%Although meta-learner ensemble models proposed by \cite{lightweight} is robust enough to effectively detect malware, it is not compatible with low-end AIoT devices due to the library usage limitation. Hence, this research aims to optimize and make the super learner model become more enery-friendly and compatible for low-end AIoT devices by reducing not only the memory usage, but also minimizing the library dependencies.

\section{Optimizing Meta-Learner Ensemble Model}
\label{sec:Optimization}
This research focuses on optimizing the super learner model, as it exhibited the best overall performance metrics (accuracy, FPR, and inference duration) per \cite{lightweight}. Our optimization maintains the same super learner structure, as shown in Table~\ref{tbl:layer constituent}. 
    
We employ three optimization techniques to enable more devices to deploy ensemble ML malware detection. First, we convert the original ML model by minimizing Python ML library usage, i.e., minimizing the overhead associated with the ML model. The rationale is that libraries reserve memory for data structures regardless of use and tend to have more functions than necessary. Then, we reduce the ML model features in an accuracy loss-minimizing manner, i.e., we reduce the number of decision points and model size. Finally, we re-code the ML models into C (C addresses memory management better and more flexibly than Python.)

\subsection{Model Conversion}
    
    In order to remove the ML model's library dependencies, each trained model needs to be converted to code free of additional libraries. In this research, two popular Python libraries are used for accomplishing this task, M2CGEN\cite{m2cgen} and emlearn\cite{emlearn}.

    \subsubsection{M2CGEN}

    M2CGEN is a Python library that can convert a subset of the scikit\cite{sklearn_api} models to code that does not require additional libraries. The advantage of using M2CGEN compared to other conversion libraries is that the vast array of scikit ML model conversions is supported, and the resultant conversion is based on the probability prediction function of scikit models, which is essential for the super learner. Unfortunately, MLPs are not supported by M2CGEN. Hence, M2CGEN is used in this research to convert the proposed super learner's RF, logistic regression (LR), and DT models.
    
    \subsubsection{emlearn}
    
    emlearn provides similar functionality to M2CGEN. For the model to run in C, emlearn provides additional header files to be included in the source file. One disadvantage of using emlearn is that for most of the classification ML models it supports, except MLP, the output given is the class to which the data belongs instead of the probability of belonging to each class. For MLP, emlearn can output which class the data belongs to and the probability of belonging to each class. Hence, in this research, emlearn is used to convert the MLP model of the proposed super learner.

\subsection{Model Optimization}
    Optimization is done to reduce model redundancy so that the memory usage of the super learner model can be decreased. In this research, two techniques are employed. The first technique reduces the number of trees in the RF model, while the second removes negligible nodes in the MLP models.

    \subsubsection{Random Forest tree trimming}
    After converting the RF using M2CGEN, it was found that there are identical DT in the RF model. The identical DTs increase memory usage yet do not contribute substantially toward the accuracy of the RF. In this research, because the RF is used in the base layer, it is assumed that the LR and the models in the subsequent layers can compensate for the reduction in accuracy. Hence, the number of trees in the RF can be decreased, after which the whole super learner model can be retrained to fit the smaller RF model better.

    \subsubsection{Removing negligible nodes in Multi-Layer Perceptrons}
    After extracting the weights from the MLP model, it was found that there were multiple nodes considered as having negligible contribution, i.e., where all of its input and output weights are 0. In this research, the negligible contribution nodes are manually removed to reduce memory usage and running time. After deleting the negligible nodes in MLP models in the intermediate layer and meta layer, the super learner model is not retrained because the deletion of negligible nodes does not appreciably affect the model's performance.

\subsection{Model Feature Reduction}
\begin{figure}[t]
        \centering
        \includegraphics[width = \columnwidth]{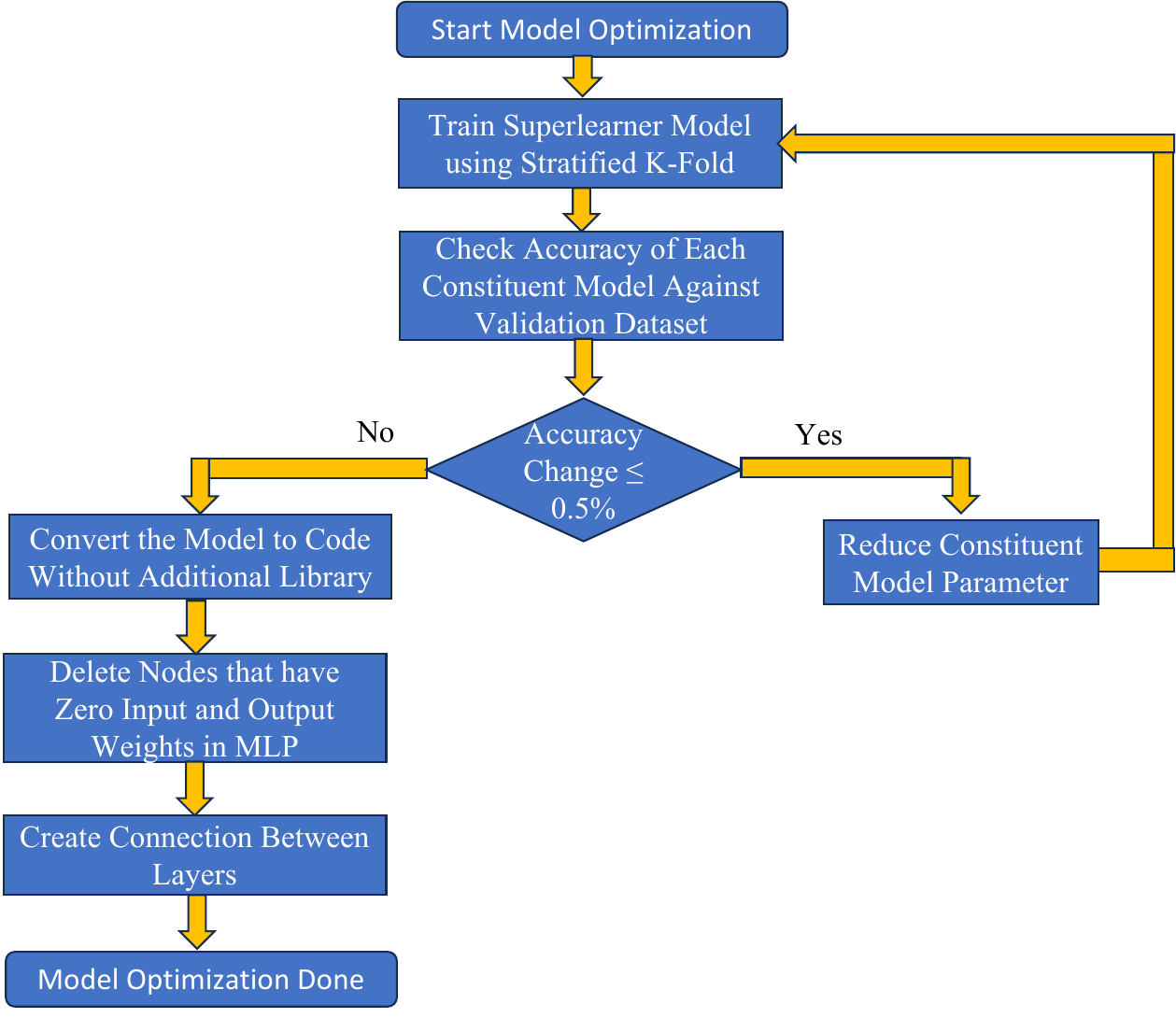}
        \caption{Model Feature Reduction workflow. The ensemble model is composed of constituent weak learners. Only when an optimization on the RF produces an accuracy change $\le 0.5\%$ is the previous model updated with the optimized model. The previous model is retained if an optimization results in an accuracy change $>0.5\%$. After which, the MLPs are optimized.} %The workflow of the solution proposed}
        \label{fig:Model Conversion Figure}
    \end{figure}

    Figure~\ref{fig:Model Conversion Figure} shows the conversion workflow proposed for model feature reduction.  
    In the first step, we train the super learner model. Then, in the second step, we check the model's accuracy. In the third step, if there was no previous model, or the current model was within $0.5\%$ of the accuracy of the previous model, we store our current model and, in the fourth step, proceed to reduce a constituent RF or MLP model's parameters, then we return to the first step.
    If the current model, post-parameter reduction, suffers from a greater than 0.5\% accuracy loss, then in the fifth step, we use our previous model and convert it to a library-free version. After which, in the sixth step, we optimize the model by removing nodes with zero-weight inputs and outputs from the MLP. In the final step, we connect the layers and finalize the reduced ML model.
    
    We reduced the DTs in the RF by multiples of 10. The optimization of the model results in the reduction of the number of DTs in the RF from 40 trees to 10 trees. The optimization of the MLP results in a reduction of the nodes in the hidden layers of the MLP model in the intermediate layer from (5, 5) to (1, 2) and the meta layer from (12, 12) to (8, 8). Note: (1, 2) means (1 node for the first layer and 2 nodes for the second layer). %so 1 is 1st hidden, 2 is the 2nd hidden
    %\textcolor{red}{Explain what this means}
    
\section{Simulation}
    \label{sec:Simulation}
\subsection{ML Setup}
    Our research compared a Raspberry Pi 4 (4 GB) model B against a Google Colab simulation of our ML models.  The same Google Colab session, configured with an Intel\textsuperscript{\textregistered} Xeon\textsuperscript{\textregistered} processor running at a 2.20 GHz clock rate, 13 GB RAM, and 100 GB storage configuration, was used for performance comparisons during optimization.
    
\subsection{Dataset}
\begin{figure}[t]
        \centering
        \includegraphics[width = \columnwidth]{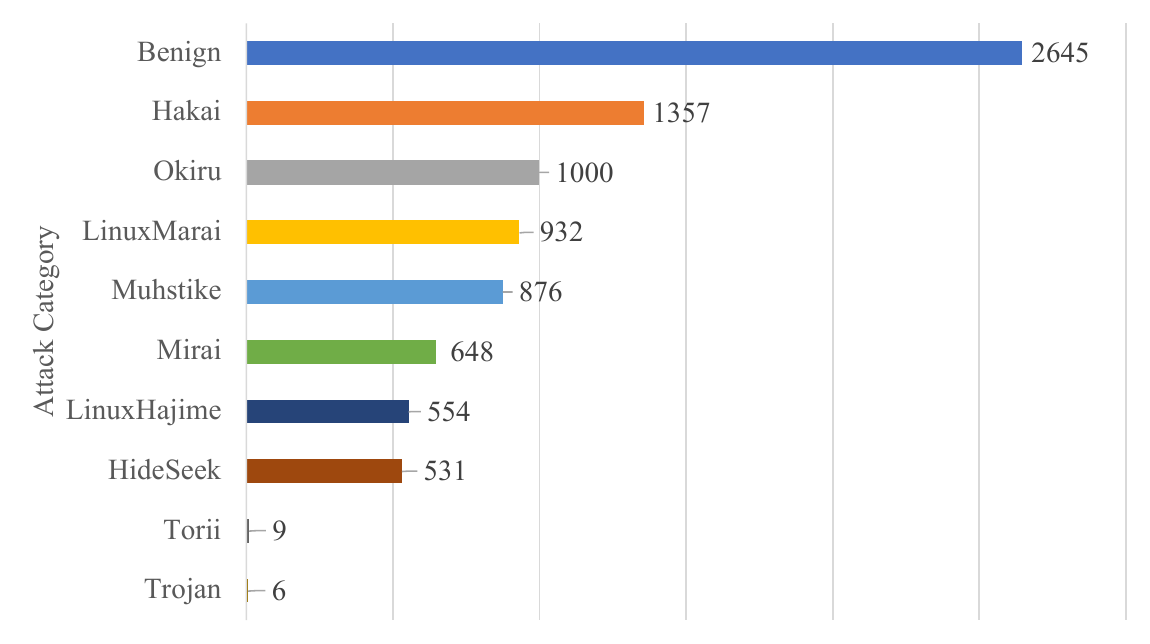}
        \caption{The attack type distribution of the Aposemat of IoT-23 dataset}
        \label{fig:Aposemat Data Dist}
    \end{figure}

    For this research, we used the Aposemat IoT-23 dataset \cite{Aposemat} to train and validate the super learner model. The dataset includes 20 IoT malware traffic captures of 5,931 flows and three benign traffic captures of 2,645 flows. The malware distribution in the dataset is shown in Figure~\ref{fig:Aposemat Data Dist}, and the statistical features used in this research are shown in Table~\ref{tbl:feature_used}. Our model used a training-to-validation dataset ratio of 8:2 with the same ratio of malware and benign data in each group (stratified and split based on the label).
    
    \begin{table}[b]
    \centering
    \caption{Features of the Aposemat IoT-23 dataset used for this research}
    \label{tbl:feature_used}
    %\resizebox{\columnwidth}{!}{%
    \begin{tabular}{|c|c|}
    \hline
    \textbf{Feature Name} & \textbf{Description}                         \\ \hline
    id.orig\_h            & Server IP address                            \\ \hline
    id.orig\_p            & Server port                                  \\ \hline
    id.resp\_h            & Client IP address                            \\ \hline
    id.resp\_p            & Client port                                  \\ \hline
    proto                 & Transaction protocol                         \\ \hline
    service               & http, ftp, smtp, ssh, dns, etc.              \\ \hline
    duration              & Total flow duration                          \\ \hline
    orig\_bytes           & Transaction bytes from server to client      \\ \hline
    resp\_bytes           & Transaction bytes from client to server      \\ \hline
    conn\_state           & Connection state                             \\ \hline
    history               & The history of the packets sent              \\ \hline
    orig\_pkts            & Number of packets sent by server to client   \\ \hline
    orig\_ip\_bytes       & Flow of bytes from server to client          \\ \hline
    resp\_pkts            & Number of packets sent by client to server   \\ \hline
    resp\_ip\_bytes       & Flow of bytes from client to server          \\ \hline
    \end{tabular}%
    %}
    \end{table}
    
\subsection{Training}

    The MLENS library uses scikit models to construct the meta-learner ensemble model. Unfortunately, the trained models cannot be directly extracted from MLENS because, during the training of the meta-learner ensemble model, MLENS makes a copy of the scikit models declared to increase the training speed. Hence, the training of the super learner model derived in this research is done manually using scikit models and reverse engineering the MLENS library to construct an equivalent ensemble model.

\begin{figure}[t]
        \centering
        \includegraphics[width = \columnwidth]{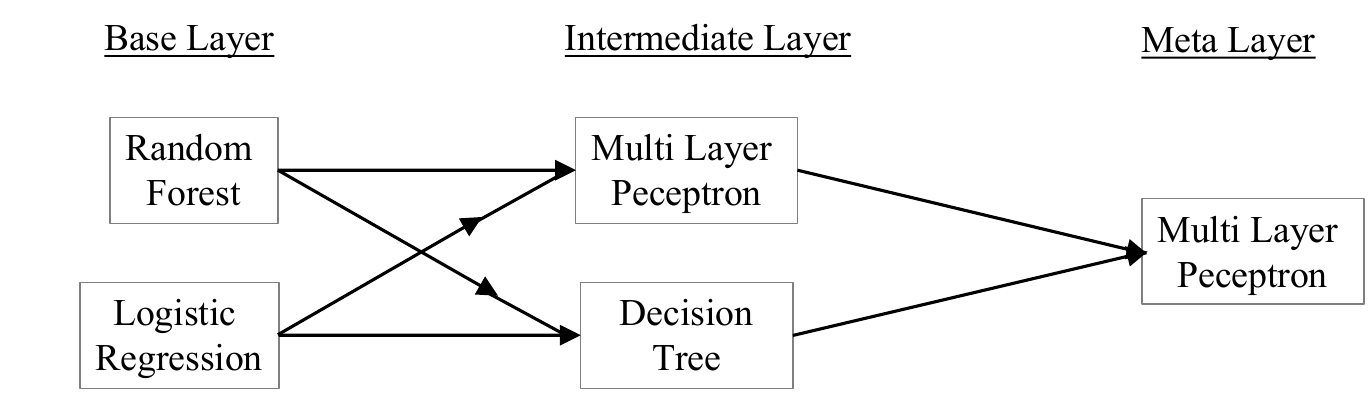}
        \caption{The super learner model scheme used in this research based on \cite{lightweight}.}
        \label{fig:super learner Scheme}
    \end{figure}

    Based on the source code of MLENS, each layer uses the prediction probability result of the immediately previous layer as the input, with the base layer using the dataset as the input, the intermediate layer using the base layer's predictions as input, and the meta (final) layer using the second layer's predictions as input. The resulting meta-learning ensemble model scheme is shown in Figure~\ref{fig:super learner Scheme}.
        
    In this research, the training of the weak learners involves the usage of stratified K-fold sampling in order to make the training input of the models in the next layer based on the prediction results of the previous layer; the benefit of this training topology is that it makes the weak learners in each layer less susceptible to unknown data. The difference between stratified K-fold versus regular K-fold is located in the data distribution used in each group. Stratified K-fold groups the data such that the ratio of the classes in each group is the same, while regular K-fold does not. In this research, the k-value for the intermediate layer to meta layer is $3$ while the base layer to intermediate layer uses the k-value $2$.

\section{Discussion}
    \label{sec:Results_Discussion}
    
\begin{figure}[t]
        \centering
        \includegraphics[width = \columnwidth]{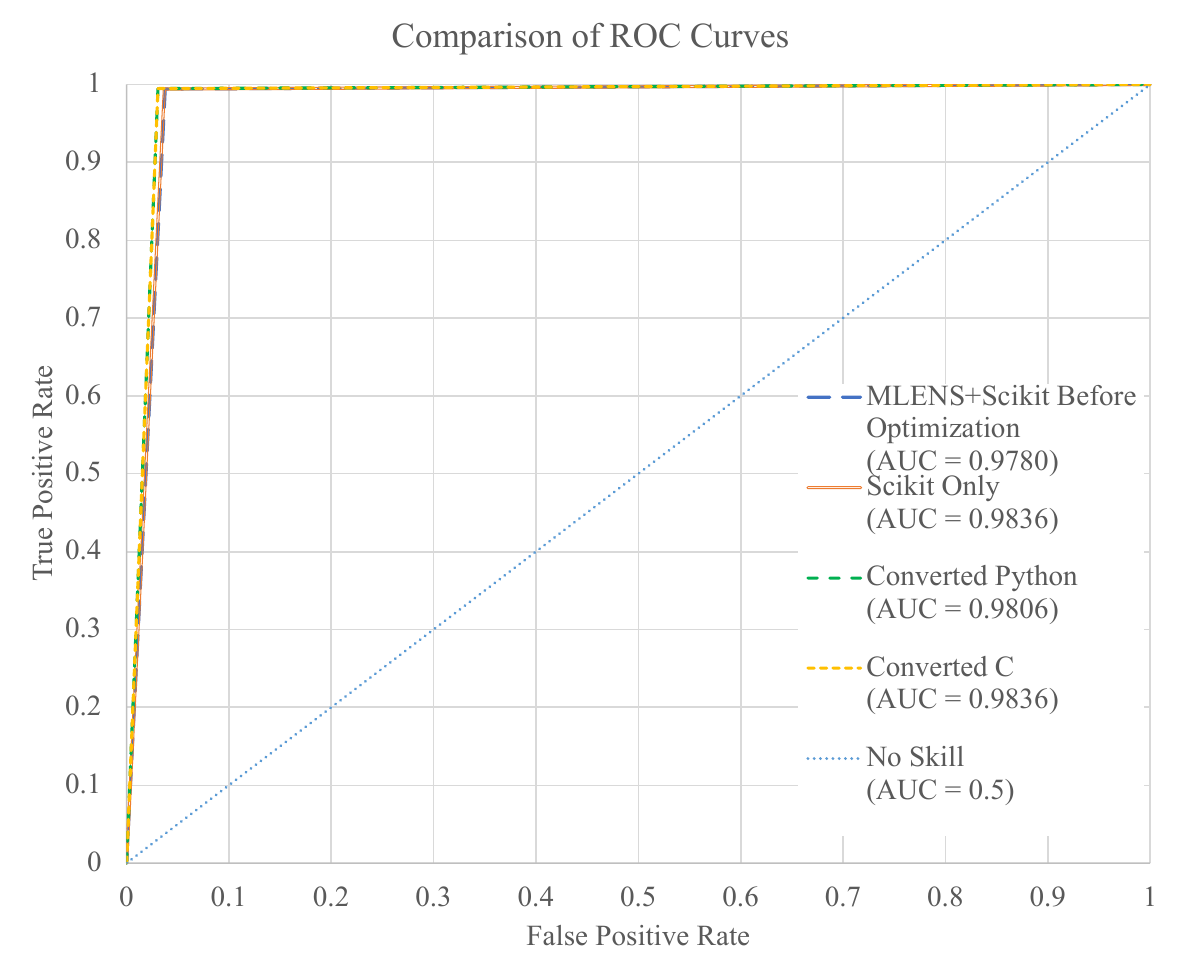}
        \caption{Model ROC detection curves}
        \label{fig:ROC AUC}
    \end{figure}

    The Receiver Operating Characteristic (ROC) curves for the models tested in Google colab are shown in Figure~\ref{fig:ROC AUC}.  
    The higher the Area Under the Curve (AUC), the better a model can differentiate between true positive versus false positive cases.  The ROC curves are ordered per the optimization steps:  The first curve represents the Meta-learning ensemble model from \cite{lightweight} and is shown as the "MLENS + Scikit Before Optimization" with an AUC of 0.9780.  The second curve shows the result due to the removal of the MLENS library; it has an AUC of 0.9836.  The third curve shows the results of model optimization (RF tree trimming and MLP negligible node removal) with no additional libraries, i.e. Python only; this curve has an AUC of 0.9806.  The fourth curve shows the result of converting Python to C; the same weights were used in both models; this curve has an AUC of 0.9836.  For reference, the fifth curve shows no skill classification; this curve has an AUC of 0.5.  As can be seen, all optimized models have similar ROC performance characteristics to the original model.

    Table~\ref{tbl:reducing tree} shows the impact of tree reduction in the RF models.  As the number of trees are reduced, the inference duration decreases.  Although reducing the number of trees also reduces accuracy and lowers the ROC AUC for the RF model, the accuracy and ROC AUC of the super learner model are not affected similarly.  Thus, we can infer that the LR model partly addresses the reduction of the RF model.

    In Table~\ref{tbl:Optimization Effect}, we compare the impact of each optimization method (RF optimization and MLP optimization) on the super learner model's inference duration after library removal, yet prior to C conversion.  We first notice that, compared to RF tree trimming, removing negligible MLP nodes has a more pronounced impact on the inference duration of the super learner model.  Trimming the RF trees, shown as (A), reduces the inference duration by an average of 30 seconds.  Separately, we observe that removing negligible MLP nodes, shown as (B), reduces the inference duration by approximately 60 seconds, on average.  The combined effects, on average, result in an inference duration reduction of approximately 90 seconds, i.e., the optimization reduced our ensemble model's inference duration by $\approx 51.4\%$.
    
    %in Table~\ref{tbl:Conversion Effect} 
    %Table~\ref{tbl:reducing tree} and Table~\ref{tbl:Optimization Effect} shows the effect of optimization. Table~\ref{tbl:reducing tree} shows the effect of reducing the number of trees in the random forest with respect to the random forest model's and super learner model's accuracy and ROC AUC, while Table~\ref{tbl:Optimization Effect} shows the effect of each optimization methods proposed in section V, with respect to the prediction duration, after the conversion to python language. 

    %\begin{minipage}[b]{1.0\columnwidth}
    
    \begin{table}[b]
    \centering
    \caption{Effect of reducing number of trees in random forest, run on Google Colab session}
    \label{tbl:reducing tree}
    %\resizebox{\columnwidth}{!}{%
    \begin{tabular}{|c|c|cc|cc|}
    \hline
    \multirow{3}{*}{\begin{tabular}[c]{@{}c@{}}\textbf{Number} \\ \textbf{of Trees}\end{tabular}} & \multirow{3}{*}{\begin{tabular}[c]{@{}c@{}}\textbf{Inference} \\ \textbf{Duration}\end{tabular}} & \multicolumn{2}{c|}{\textbf{Random Forest}} & \multicolumn{2}{c|}{\textbf{Super Learner}} \\ \cline{3-6} 
   &      & \multicolumn{1}{c|}{\begin{tabular}[c]{@{}c@{}}Accuracy \\ (\%)\end{tabular}} & \begin{tabular}[c]{@{}c@{}}ROC \\ AUC (\%)\end{tabular} & \multicolumn{1}{c|}{\begin{tabular}[c]{@{}c@{}}Accuracy \\ (\%)\end{tabular}} & \begin{tabular}[c]{@{}c@{}}ROC \\ AUC (\%)\end{tabular} \\ \hline
    40 & 18.8 sec & \multicolumn{1}{c|}{99.37}    & 97.918  & \multicolumn{1}{c|}{99.34}    & 97.761  \\ \hline
    30 & 15.3 sec & \multicolumn{1}{c|}{99.23}    & 97.314  & \multicolumn{1}{c|}{99.34}    & 97.849  \\ \hline    
    20 & 14.8 sec & \multicolumn{1}{c|}{99.21}    & 97.289  & \multicolumn{1}{c|}{98.99}    & 97.178  \\ \hline    
    10 & 13.3 sec & \multicolumn{1}{c|}{98.76}    & 97.221  & \multicolumn{1}{c|}{99.48}    & 98.364  \\ \hline
    \end{tabular}%
    %}
    \end{table}

    \begin{table}[b]
    \centering
    \caption{Optimization effect in converted python model, run on Google Colab session}
    \label{tbl:Optimization Effect}
    %\resizebox{\columnwidth}{!}{%
    \begin{tabular}{|c|c|} %c|}
    \hline
    \multicolumn{1}{|c|}{\textbf{Model Reduction Step}}               & \makecell{\textbf{Inference} \\ \textbf{Duration}} \\ \hline% & \makecell{Standard \\ Deviation} \\ \hline
    Initial                              & 2 min 55 sec \\ \hline% & 6.46 sec           \\ \hline
    Reducing tree in RF (A)              & 2 min 25 sec \\ \hline%& 4.55 sec           \\ \hline
    Deleting negligible nodes of MLP (B) & 1 min 56 sec \\ \hline%& 2.71 sec           \\ \hline
    (A) + (B)                            & 1 min 25 sec \\ \hline%& 2.25 sec           \\ \hline
    \end{tabular}%
    %}
    \end{table}

    %\end{minipage}

    %Besides comparing the effect of optimization, the effect of conversion is also tested, which is shown in Table~\ref{tbl:Conversion Effect}. 
    
    \begin{table}[b]
    \centering
    \caption{ML model conversion performance metrics (Raspberry Pi)}
    \label{tbl:Conversion Effect}
    %\resizebox{\columnwidth}{!}{%
    \begin{tabular}{|c|c|c|c|c|} %c|} %{|S|P|P|A|A|A|A|P|}
    \hline
    \textbf{Super Learner Model} & \makecell{Inference \\ Duration} & \makecell{Accuracy \\ (\%)} & \makecell{TPR \\ (\%)} & \makecell{FPR \\ (\%)} \\ \hline %& \makecell{Throughput \\ (kBps)} \\ \hline
    \makecell{MLENS + Scikit} %& % \\ Pre-Optimization} 
    %& 25.8 sec  & 99.40 & 99.40 & 3.80 \\ \hline %& 131.40 
    & 5.46 sec  & 99.40 & 99.40 & 3.80 \\ \hline %& 131.40 
    \makecell{Scikit} %& % \\ Post-Optimization}    
    %& 9.17 sec  & 99.48 & 99.80 & 3.08 \\ \hline %& 283.86 \\ \hline
    & 7.79 sec  & 99.48 & 99.80 & 3.08 \\ \hline %& 283.86 \\ \hline
    \makecell{Python (no ML library)} %&  \\ Post-Optimization}         
    %& 229 sec   & 99.41 & 99.81 & 3.69 \\ \hline %& 11.37  \\ \hline
    & 229 sec   & 99.41 & 99.81 & 3.69 \\ \hline %& 11.37  \\ \hline
    \makecell{C (conversion)} %& % \\ Post-Optimization}              
    & 9.57 sec & 99.48 & 99.80 & 3.08 \\ \hline %& 272.14 \\ \hline %9.565
    \end{tabular}%
    %}
    \end{table}
    
    \begin{table}[b]
    \centering
    \caption{ML model conversion memory requirements (Raspberry Pi)}
    \label{tbl:Conversion Effect Mem}
    %\resizebox{\columnwidth}{!}{%
    \begin{tabular}{|c|c|c|} %{|S|P|P|A|A|A|A|P|}
    \hline
    \textbf{Super Learner Model} & \makecell{Memory \\ Usage (MB)} & \makecell{Model Size \\ (kB)} \\ \hline
    \makecell{MLENS + scikit} %& % Pre-Optimization} 
    %& 1238.66 & 563.7 \\ \hline
    & 1095 & 92 \\ \hline
    \makecell{scikit} %& % \\ Post-Optimization}   
    %& 1344.83 & 39.96 \\ \hline
    & 1382 & 48 \\ \hline
    \makecell{Python (no ML library)} %& % \\ Post-Optimization}        
    %& 1723.06 & 24.58 \\ \hline
    & 1770 & 28 \\ \hline
    \makecell{C (conversion)} %& % \\ Post-Optimization}             
    %& 0.709   & 38.72 \\ \hline
    & 3.4   & 37.6 \\ \hline
    \end{tabular}%
    %}
    \end{table}

    In Table~\ref{tbl:Conversion Effect}, we show how the optimization process affects the super learner model in terms of inference performance metrics (inference duration, accuracy, True Positive Rate (TPR), and FPR) as measured on a Raspberry Pi 4 Model B (4GB model). While we observe that the conversion from the scikit ML model to C does not appreciably affect the model's detection performance, shown by the same ROC AUC score, accuracy, TPR, and FPR, we observe the interference duration metric changes. Removing MLENS increases the inference duration by $\approx 42.7\%$, thus indicating that removing MLENS impacts the ML model's performance negatively. We also observe that using pure Python increases the inference duration by $\approx 29.4\times$, relative to scikit. We can thus infer that scikit manages data structures and ML model construction better than a pure Python implementation; this depends on how the pure Python version is implemented. Finally, we observe in the post-C, ML library-free conversion that the inference duration is similar to the scikit-only Python implementation.
    One possible explanation for this phenomenon is the difference in the coding language's memory management and that Python is interpreted and then executed as a low-level language like C. 
    % Python processes memory allocation differently than C, i.e., freeing memory does not release it from Python control; instead, it is kept for future use as free memory.
    While a direct C implementation benefits from better memory allocation and control, manual allocation is necessary to improve performance. In comparison, Python focuses on ease of use and does not approach memory allocation as a major concern.

    Similarly, in Table~\ref{tbl:Conversion Effect Mem}, we show how the optimization process affects the super learner model in terms of inference memory (RAM + model size\footnote{Pickle is used for Python model size comparison. For C, source and header files sizes are summed.}) requirements. % as measured on a Raspberry Pi 4 Model B (4GB model). 
    The base performance metrics, Pre-Optimization, shown as MLENS + scikit, demonstrate the reference no-conversion metrics. We underscore here a major issue precluding the use of this incarnation of the super learner model: The memory required for the run time environment is over 1000 MB, %2)  The model size is over 560kB. 
    Low-end IoT devices are unlikely to support such a high demand for run time resources% or, for that matter, fit the ML model into memory
    \footnote{Although Python memory resources can have a limit imposed, limiting memory may prevent Python from loading libraries and/or increase the inference duration.}\textsuperscript{,}\footnote{Python does not release memory back to the system when data structures are destroyed, per se, rather garbage collection is used. The actual release of memory back to the Operating System can occur later in time.}. As we reduce library dependencies by removing MLENS, shown as scikit, we notice a couple of interesting metric changes. Removing MLENS increases the inference duration by $\approx42.7\%$ while maintaining similar accuracy, TPR, and FPR metrics. % \textcolor{red}{should explain FPR difference?}. 
    Despite the reduction in model size by $\approx47.8\%$, the run-time memory requirements still far exceed the resources available on low-end IoT devices. After all ML libraries are removed, shown as Python (no ML library), we observe that the ML model size reduces by $\approx41.7\%$ of the scikit-only model size, yet unfortunately, removing the libraries has resulted in the RAM memory requirements to increase by $\approx28.1\%$, still leaving the resulting design untenable for low-end IoT devices. % 
    One possible explanation for the drastic changes in memory requirements is the usage of an ineffective MLP dense algorithm in Python referenced from \cite{ann_python}, which is used in this research due to the ease of inputting the extracted weights into the algorithm directly.
    % Check (Jun 22)
    %I think that there is a possibility the one in Python calculates the memory usage as the whole data is run at the same time. I know for sure that for the one in C, the memory usage is calculated as the peak memory usage as the data is being processed one by one, and the resultant was appended together as a matrix.
    %
    Finally, we examine the C version, shown as C (conversion) of the ML model (C handles memory management better than Python.)  We observe that although the model size has increased relative to the pure Python version by $\approx34.3\%$, it still fits in low-end devices. Interestingly, the Run time environment required for the C version is over $\approx95.8\%$ smaller compared to Python incarnations of the ML model. Thus, we see that our optimized C version of the super learner ML model can indeed fit inside low-end IoT devices.
    %
    
    %also included the result of the model from \cite{lightweight} as MLENS + scikit Pre-Optimization. From Table~\ref{tbl:Conversion Effect}, it is apparent that the model optimization reduced the prediction duration, which is shown by scikit and C Post-Optimization, resulting in around 2.8 times increased throughput. Although have been optimized, the Python Post-Optimization has the longest prediction duration within the models compared. One possible explanation for this phenomena is the usage of ineffective MLP dense algorithm in python referenced from \cite{ann_python}, which is used in this research due to the ease of inputting the extracted weights into the algorithm directly.
    
    %From Table~\ref{tbl:Conversion Effect} and Figure~\ref{fig:ROC AUC}, the conversion from scikit model to C seems to not affect the model's performance, shown by the same ROC AUC score, accuracy, TPR (True Positive Rate), and FPR (False Positive Rate) of the models. Although the model's performance is not affected, but the memory usage is reduced by around 1800 times when the model has been converted to C. One possible explanation for this phenomena is due to the difference in coding language memory processing. C coding language has a better memory allocation, but manual allocation is needed. While python focuses on the ease of use and the memory allocation is not python's main concern.

\section{Conclusion and Future Work}
    \label{sec:Conclusion}
    From the results shown, the optimized C implementation of the super learner ML model performs best regarding memory usage while maintaining a low inference duration with similar detection metrics. Although this research focuses on optimizing the super learner model for lower-end AIoT devices, our methodology is predicted to yield a similar effect when used with other ensemble models.
    
    In the future, we plan to test the proposed model feature reduction workflow on other AI models and datasets.  In addition, we plan to test the resulting C model directly on low-end AIoT devices; we expect it will run on the ESP32 with external RAM based on current memory usage. 

\FloatBarrier

\bibliography{Ref_final}
\bibliographystyle{IEEEtran}

\end{document}